\documentclass{aa}  
\usepackage{graphicx}
\usepackage{amsmath}
\usepackage{amsfonts}
\usepackage{amssymb}
\usepackage{gensymb}
\usepackage{textcomp}
\usepackage[varg]{txfonts}
\usepackage{natbib}
\usepackage{epstopdf}

\usepackage{color}
\definecolor{red}{rgb}{1.0, 0.1, 0.1}
                                
\begin{document}

   \title{Compact solar UV burst triggered in a magnetic field with a fan-spine topology}
   \author{Authors}
   \author{L.~P.~Chitta\inst{1}, H.~Peter\inst{1}, P.~R.~Young\inst{2,}\inst{3,}\inst{4}, \and Y.-M.~Huang\inst{5,}\inst{6}}

   \institute{Max Planck Institute for Solar System Research, 37077, G\"{o}ttingen, Germany\\
              \email{chitta@mps.mpg.de} 
              \and 
              College of Science, George Mason University, 4400 
              University Drive, Fairfax, VA 22030, USA
              \and
              NASA Goddard Space Flight Center, Solar Physics Laboratory, Greenbelt, MD
              20771, USA
              \and
              Northumbria University, Newcastle Upon Tyne NE1 8ST, UK
              \and
              Max-Planck-Princeton Center for Plasma Physics, Princeton, NJ 08540, USA
              \and
              Department of Astrophysical Sciences and Princeton Plasma Physics Laboratory, Princeton University,                      Princeton, NJ 08540, USA}

   \date{Received 21 March 2017 / Accepted 24 June 2017}

 
\abstract
{Solar ultraviolet (UV) bursts are small-scale features that exhibit intermittent brightenings that are thought to be due to magnetic reconnection. They are observed abundantly in the chromosphere and transition region, in particular in active regions.}
%
{We investigate in detail a UV burst related to a magnetic feature that is advected by the moat flow from a sunspot towards a pore. The moving feature is parasitic in that its magnetic polarity is opposite to that of the spot and the pore. This comparably simple photospheric magnetic field distribution allows for an unambiguous interpretation of the magnetic geometry leading to the onset of the observed UV burst.}
%
{We used UV spectroscopic and slit-jaw observations from the Interface Region Imaging Spectrograph (IRIS) to identify and study chromospheric and transition region spectral signatures of said UV burst. To investigate the magnetic topology surrounding the UV burst, we used a two-hour-long time sequence of simultaneous line-of-sight magnetograms from the Helioseismic and Magnetic Imager (HMI) and performed data-driven 3D magnetic field extrapolations by means of a magnetofrictional relaxation technique. We can connect UV burst signatures to the overlying extreme UV (EUV) coronal loops observed by the Atmospheric Imaging Assembly (AIA).}
%
{The UV burst shows a variety of extremely broad line profiles indicating plasma flows in excess of $\pm$200\,km\,s$^{-1}$ at times. The whole structure is divided into two spatially distinct zones of predominantly up- and downflows. The magnetic field extrapolations show a persistent fan-spine magnetic topology at the UV burst. The associated 3D magnetic null point exists at a height of about 500 km above the photosphere and evolves co-spatially with the observed UV burst. The EUV emission at the footpoints of coronal loops is correlated with the evolution of the underlying UV burst.}
%
{The magnetic field around the null point is sheared by photospheric motions, triggering magnetic reconnection that ultimately powers the observed UV burst and energises the overlying coronal loops. The location of the null point suggests that the burst is triggered low in the solar chromosphere.}
  
   \titlerunning{Solar UV burst}
   \authorrunning{Chitta et al.}
   \maketitle

%

\section{Introduction\label{intr}}

Solar active regions and the quiet-Sun magnetic network host a plethora of small-scale magnetic activity that leave interesting observable signatures in the atmosphere of the Sun. These include intensity enhancements and fluctuations that are rapid compared to the background emission level, and/or complex spectral line profiles. Depending on the magnetic environment where these events occur and the type of spectral signatures they leave, the events have received a variety of names in the literature. In newly emerging active regions, Ellerman bombs in the photosphere \citep{1917ApJ....46..298E,2002ApJ...575..506G}, flaring-arch filaments and UV bursts in the chromosphere and transition region  \citep{2013JPhCS.440a2007R,2014Sci...346C.315P} are observed. In the magnetic network, explosive events \citep{1989SoPh..123...41D} and blinkers \citep{1997SoPh..175..467H} in the transition region are observed, as well as more recently, Ellerman bombs in the quiet Sun photosphere \citep{2016A&A...592A.100R}. It is widely believed that all these events are signatures of energy release that is due to magnetic reconnection, the same process that is also thought to play a key role in large-scale solar eruptions and flares.

Since the launch of the Interface Region Imaging Spectrograph \citep[IRIS;][]{2014SoPh..289.2733D}, it has become feasible to follow these small-scale transient events or bursts and their connection from the photosphere to the corona \citep{2015ApJ...812...11V,2016ApJ...824...96T}. When combined with photospheric magnetic field information, these transients appear to be resulting from the interaction of closely spaced magnetic field structures with opposite polarities. During this interaction, a cancellation in the surface magnetic flux density of one or both of the polarities is observed, and it is thought that in the process, the energy is released through magnetic reconnection that ultimately energises the burst \citep{2014Sci...346C.315P,2016MNRAS.463.2190N}. However, in some cases, the observed magnetic flux cancellation does not lead to bursty events \citep{2016MNRAS.463.2190N}. 

Numerical models of small-scale bursts and jets in the lower
solar atmosphere reproduce the observed events qualitatively by typically assuming an anti-parallel configuration of the magnetic field or a Harris current sheet \citep[e.g.][]{2015ApJ...799...79N,2015ApJ...813...86I}. Although simulations of small-scale bursts with complex magnetic geometries such as fan-spine topology and null points exist \citep[e.g.][]{2009ApJ...702....1H}, the observational basis of these geometries and the height of magnetic reconnection in the solar atmosphere, as supported by magnetic models and extrapolations of observed magnetic field, are not well established. On the other hand, to explain large-scale explosive events such as circular ribbon flares, a fan-spine magnetic topology is often invoked \citep[e.g.][]{1998ApJ...502L.181A,2009ApJ...700..559M,2013ApJ...778..139S}. 

The question is whether complex magnetic topologies also trigger bursts at small spatial scales. If this were the case, the topological similarities in the magnetic field would enable us to draw parallels between small-scale bursts and large-scale flares. High-resolution observations show that complex magnetic configurations at small spatial scales exists \citep{2017ApJS..229....4C}. Recently, \citet{2016ApJ...819L...3Z} reported observations of a resolved fan-spine structure in a chromospheric jet. However, their observations have been made close to the limb, and for this reason, a detailed study of the magnetic field topology is lacking. 

\begin{figure*}[!htb]
\centering
\includegraphics[width=17cm]{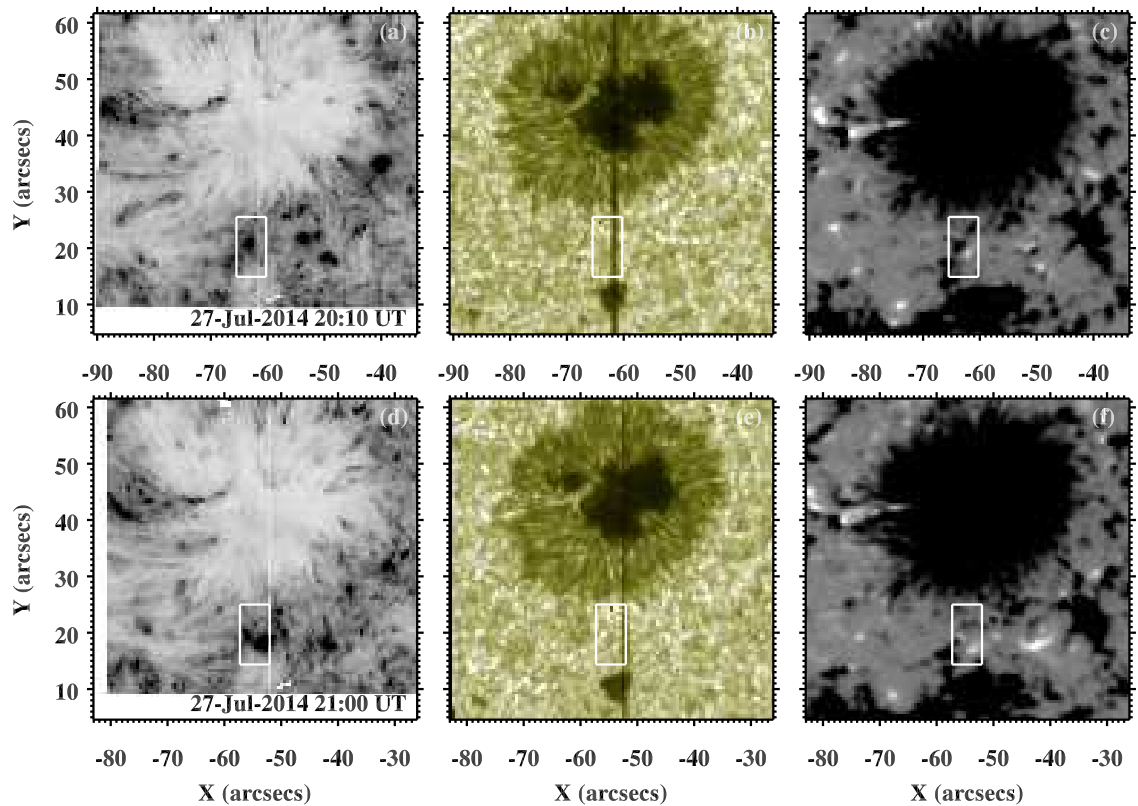}
\caption{Spatial evolution of a compact UV burst and the surrounding magnetic environment as seen with  IRIS and SDO/HMI. Panels \textbf{a)} and \textbf{d)} show two SJI 1400\,\AA\ images separated by 50\,minutes that mainly outline chromospheric and transition region features. Panels \textbf{b)} and \textbf{e)} show co-temporal SJI 2832\,\AA\ images displaying photospheric granulation. Co-temporal HMI line-of-sight magnetic field maps saturated at $\pm250$\,G are plotted in panels \textbf{c)} and \textbf{f)}. The location of the compact UV burst is outlined by a white box in all the panels. Here the white box marks the sub-field of view of raster maps we used for the
more detailed investigation. This sub-field of view is shown as an inset map in the left panel of Fig.\,\ref{spect} and in the two panels of Fig\,\ref{dopp}. The two SJI 1400\,\AA\ images are the negatives of the square-root-scaled intensity. The vertical lines in panels \textbf{b)} and \textbf{e)} show the locations of the slit positions at these times. See Sect.\,\ref{obsv2}.
\label{context}}
\end{figure*}

Here we report the observation of a UV burst near solar disk centre. Unlike typical UV bursts, the event under investigation was observed in an evolved active region where a systematic flux emergence is absent near the burst during its evolution. From magnetic field extrapolations, we identify a fan-spine topology of the magnetic field and locate the associated null point. This enables us to derive the height above the photosphere where the UV burst is assumed to have been triggered. 


\section{Observations\label{obsv}}

On 2014 July 27, the IRIS observed a sunspot-pore pair in active region AR 12121, close to the disk centre at $(-64\arcsec,37\arcsec).$ These observations consist of two-hour-long time series of slit-jaw images (SJI) from different channels and medium-dense 16-step spectroscopic raster scans, starting at 20:04\,UT. The step size of the raster maps in the direction of the scan is 0.35\arcsec, covering a full field of view of 5\arcsec$\times$64\arcsec. We used level-2 data from these observations\footnote{available at http://iris.lmsal.com/} for our analysis. The slit-jaw images are recorded with a cadence of 22\,s, while the raster maps have a cadence of about 88\,s with 4\,s exposures per step. For the spectroscopic analysis, we concentrated on Mg\,{\sc ii} and Si\,{\sc iv,} which have formation temperatures of log$T$\,[K] $=$ 4.0, and 4.9, respectively. We also used the line-of-sight magnetograms from the Helioseismic and Magnetic Imager~\citep[HMI;][]{2012SoPh..275..207S} and coronal EUV imaging observations from the Atmospheric Imaging Assembly~\cite[AIA;][]{2012SoPh..275...17L}. HMI and AIA are both on board the Solar Dynamics Observatory~\citep[SDO;][]{2012SoPh..275....3P}.

\subsection{Spatial and temporal evolution of the UV burst and the underlying magnetic field\label{obsv2}}

The UV burst analysed here is a long-lived structure exhibiting several episodes of intensity enhancements, each several minutes long. Two instances of the burst are illustrated in Fig.\,\ref{context}, where we display two SJI 1400\,\AA\ maps, 50\,minutes apart (the burst is seen in the white box in the left panels). By comparing the location of this UV burst with photospheric structures (middle panels), we note that the event is being triggered in between the two darker circular magnetic features (north: larger sunspot; south: smaller pore). The HMI magnetograms (right panels) reveal that the sunspot and pore have the same negative polarity. 

\begin{figure}[!htb]
\centering
\includegraphics[width=8.5cm]{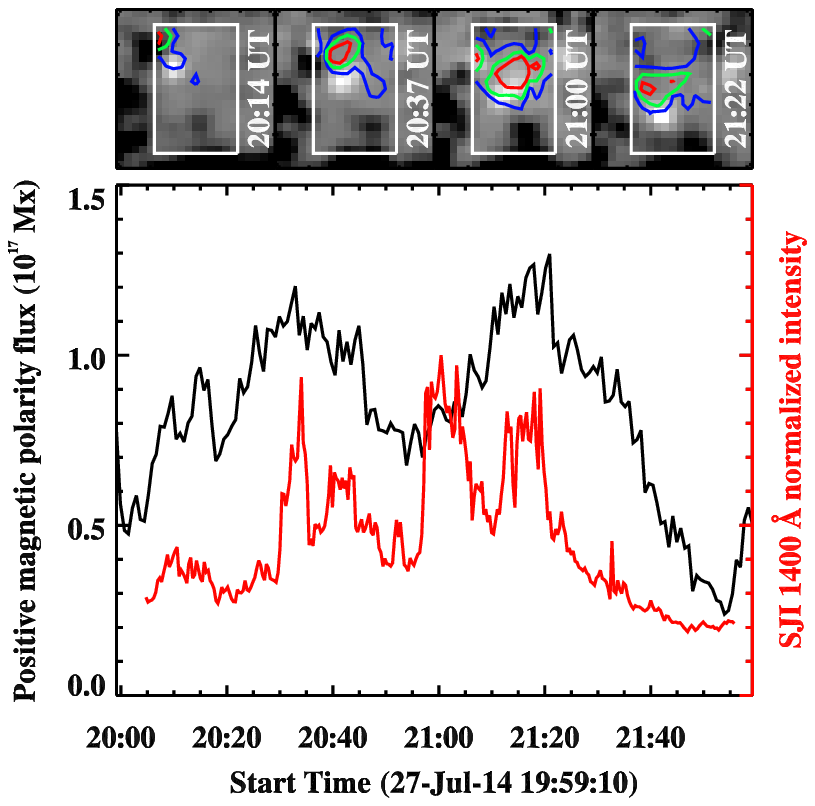}
\caption{Temporal evolution of the UV burst and the underlying magnetic field. The top panel shows a series of HMI line-of-sight magnetograms saturated at $\pm250$\,G, recorded at times as indicated in the respective tiles. The white box outlines the region with parasitic positive magnetic polarity underlying the UV burst. The location of the burst as seen in co-temporal IRIS SJI 1400\,\AA\ images is highlighted with contour lines. The blue, green, and red contours indicate SJI intensities of 100\,DN, 200\,DN, and 500\,DN, respectively. The field of view covered by each tile is marked with a smaller yellow box in Figs.\,\ref{sdo} and \ref{fan} and has an area of 10.6\arcsec$\times$10.6\arcsec. The black curve in the bottom panel is the spatially averaged magnetic flux of the positive polarity magnetic field in the region outlined by the white box (top panel). Only pixels with a magnetic flux density above 15\,G\ are considered for the averaging. The red curve is the time series of SJI intensity normalised to its maximum, showing the evolution of the UV burst (see also Fig.\,\ref{lcurves}). A two-hour-long time sequence of the magnetic maps displayed in the top panel is available as an online movie. See Sect.\,\ref{obsv2}.
\label{mflux}}
\end{figure}

A parasitic magnetic field with positive polarity is underlying the burst. As this magnetic field is advected by the moat flow of the sunspot from north to south, the UV burst is also displaced in the same direction, closely following the motion of the parasitic magnetic field. These translational motions of the magnetic field and the UV burst from north to south are demonstrated in Fig.\,\ref{mflux} (see also Fig.\,\ref{context}). The burst itself is triggered by a continual interaction of this parasitic positive-polarity field with the sunspot and pore magnetic field, including other nearby small-scale negative-polarity magnetic elements.

The temporal evolution of the average magnetic flux of the parasitic magnetic field has a timescale of about one\,hour. During the
two hours of the observations, the positive magnetic polarity flux shows a fluctuation of about 50\% or more. Towards the end of the time series, the flux of the parasitic magnetic field underlying the burst has noticeably decreased. This decrease is not simply due to the advection of the magnetic flux out of the box we considered, but is due to its cancellation with the pore (black curve, bottom panel in Fig.\,\ref{mflux}, see online movie). This is expected because the magnetic field there is being advected by the moat flow towards an opposite-polarity pore.

The intensity of the UV burst from SJI 1400\,\AA\ images shows variation at shorter timescales of 5--10\,minutes with one-minute fluctuations superimposed. There is a rapid enhancement and decline in the intensity (red curve, bottom panel in Fig.\,\ref{mflux}). However, these intensity variations of the burst do not always correlate with the cancellation of the parasitic magnetic field. This is even clearer towards the end of the time series when the magnetic flux of the parasitic element declines through flux cancellation with the pore. This cancellation does not lead to additional UV bursts, on the contrary, the burst completely disappears. Typically, photospheric flux cancellation leads to bursty signatures \citep[e.g.][]{2014Sci...346C.315P,2016MNRAS.463.2190N}. For the case analysed here, however, we find that the burst can be triggered without photospheric flux cancellation, and moreover, flux cancellation does not necessarily lead to the formation of the UV burst \citep[see also][]{2016MNRAS.463.2190N}.

\subsection{Spectral evolution\label{obsv3}}

\begin{figure*}[!htb]
\centering
\includegraphics[width=17cm]{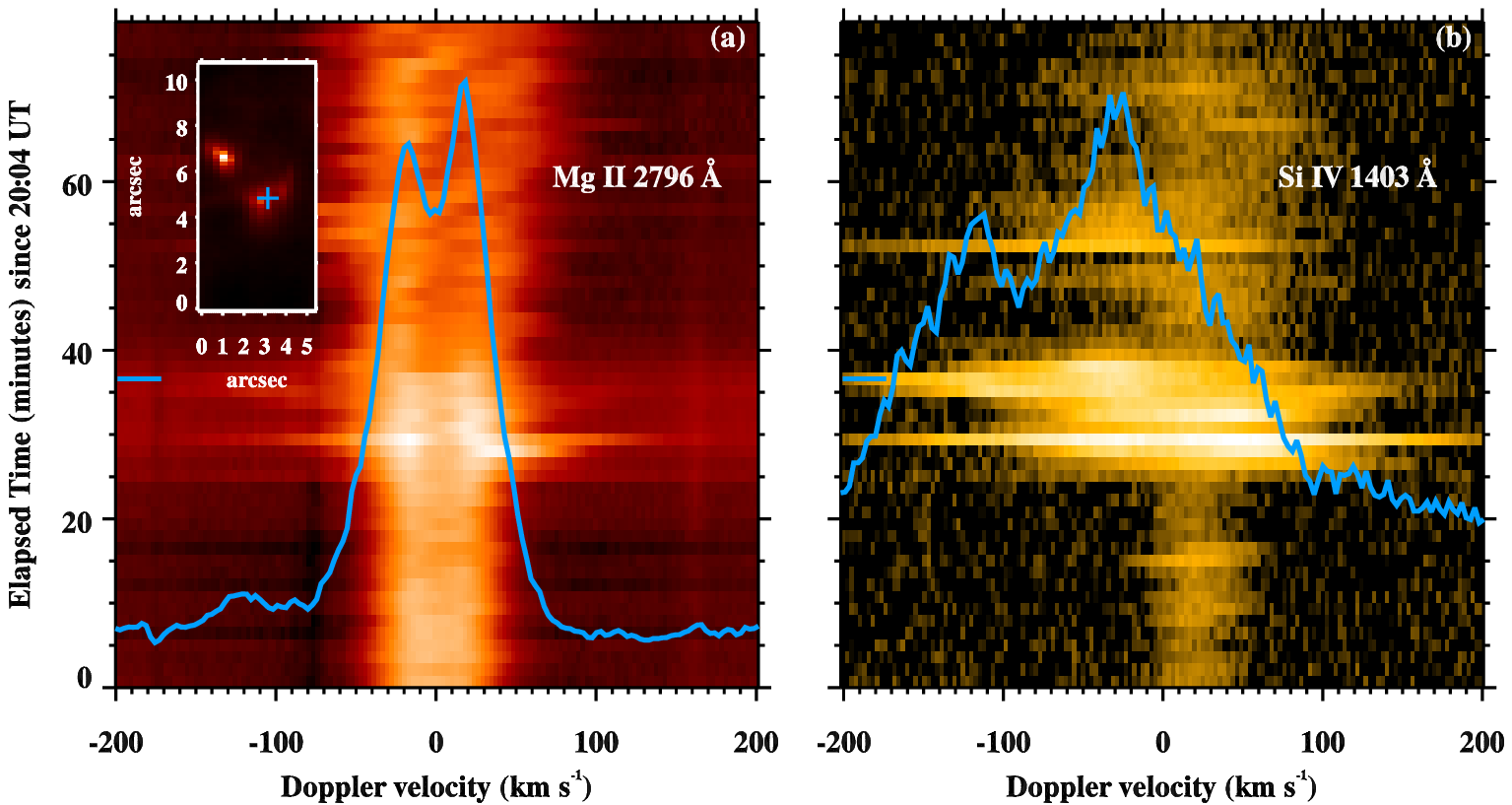}
\caption{Samples of spectral profiles from the solar chromosphere and transition region at the site of the UV burst. In panel \textbf{a)}, the background image is a Mg\,{\sc ii} $\lambda-t$ plot (in Doppler units) near 2796\,\AA, obtained from the spatial location marked by a plus in the inset. Here the inset shows a sub-field of view of the raster map at the mark of 37\,minutes (indicated by a small horizontal blue bar). The blue curve is the Mg\,{\sc ii} line profile at that instance. Panel \textbf{b)} is the same as panel \textbf{a),} but plotted for the Si\,{\sc iv} 1403\,\AA\ line that traces the solar transition region. See Sect.\,\ref{obsv3}.
\label{spect}}
\end{figure*}

This UV burst is associated with complex and broad line profiles in the chromosphere and transition region. Two representative spectral profiles each from the chromosphere and transition region from the burst are plotted in Fig.\,\ref{spect}. Here and in the rest of the paper we follow the convention that negative Doppler velocities correspond to blueshifts. In these profiles for the faster blueshifted component of Si\,{\sc iv} at $-120$\,km\,s$^{-1}$, a corresponding enhancement in the blue wing of Mg\,{\sc ii} can be seen. This suggests that the UV burst has an impact on the plasma at a broad range of temperatures and heights. Furthermore, the two blueshifted components of the Si\,{\sc iv} line in the same spatial location indicate a complex unresolved flow pattern. A series of broad line profiles can be seen between 25--40\,minutes in the background $\lambda-t$ images in the two panels at the same spatial location.

There is little or weak emission observed in the O\,{\sc iv} multiplet near 1400\,\AA, with a formation temperature of log$T$\,[K]$=$5.15, at the site of the burst. This little emission in the O\,{\sc iv} lines is consistent with recent observations of IRIS UV bursts \citep{2014Sci...346C.315P,2016ApJ...824...96T}. Typically, the absence of these O\,{\sc iv} lines is attributed to the burst being formed in a high-density atmosphere, where the O\,{\sc iv} line formation is suppressed as a result of collisional de-excitation \citep[see e.g.][cf. \citealt{1978A&A....65..215F,2015arXiv150905011Y}]{2014Sci...346C.315P}. On the other hand, we do not observe a persistent signature of the Ni\,{\sc ii} absorption line at 1393.3\,\AA, which is also the case for a few examples discussed in \citet{2016ApJ...824...96T}. This suggests that the UV burst analysed here is formed not so deep in the atmosphere but at the same time formed at heights with a sufficiently high density to suppress the formation of the O\,{\sc iv} lines.

\subsection{Spatial structure of spectral profiles and Doppler shifts\label{obsv4}}

\begin{figure}[!htb]
\centering
\includegraphics[width=8.5cm]{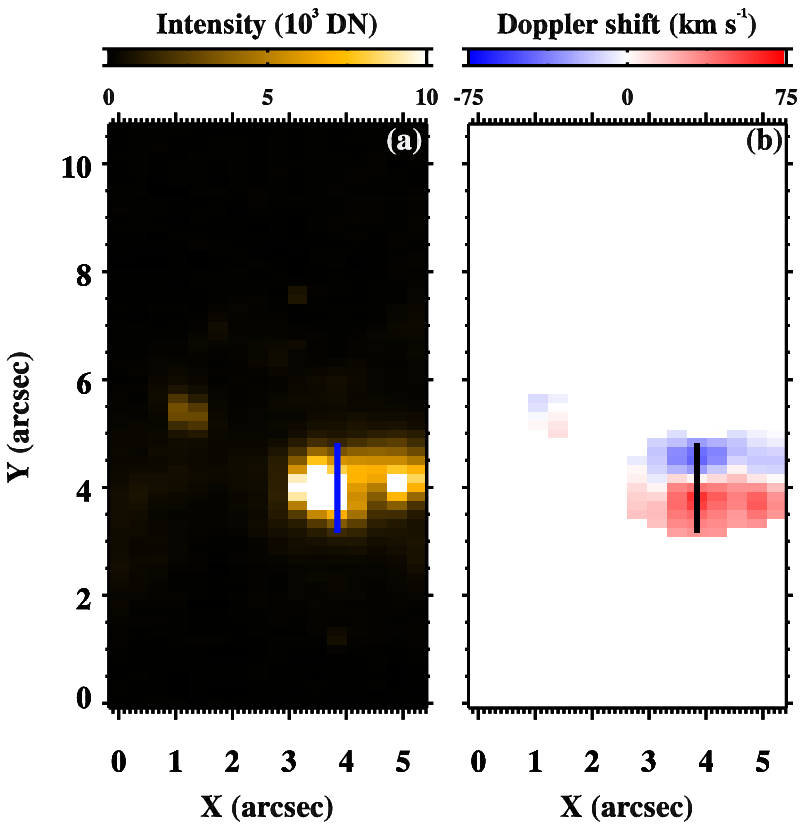}
\caption{Maps of the transition region intensity and Doppler shifts from the Si\,{\sc iv} 1403\,\AA\ line, obtained at 21:02 UT, associated with the compact UV burst. Panel \textbf{a)} shows
the spatial distribution of the Si\,{\sc iv} 1403\,\AA\ line integrated intensity within $\pm200$\,km\,s$^{-1}$ from its zero-Doppler shift. Panel \textbf{b}) shows the map of an intensity-weighted Doppler shift of the same spectral line, saturated at $\pm75$\,km\,s$^{-1}$. The vertical lines in the panels show the location along which several spectral profiles are displayed in Fig.\,\ref{avgdopp}. The temporal evolution of the intensity and Doppler-shift maps is available as an online movie. See Sect.\,\ref{obsv4}.
\label{dopp}}
\end{figure}

As we have a full spatial coverage of the burst from the successive raster maps, we exploited these data to investigate the complex spectral profiles from the burst. To this end, we computed the integrated intensity of Si\,{\sc iv} 1403\,\AA\ in the range of wavelength shifts corresponding to $\pm200$\,km\,s$^{-1}$. These integrated intensity maps reveal that the UV burst is an intense compact structure with a  spatial extension of $\approx2\arcsec$ in each direction. One of these integrated intensity maps is plotted in Fig.\,\ref{dopp}(a). The Doppler shifts provide information on the plasma flows (blueshifts correspond to upflows and redshifts to downflows). Because the spectral lines are extremely broad (for an example, see Fig.\,\ref{spect}b), we adopted the intensity-weighted spectral line shift, that is, the first moment of the line profile, to characterize the Doppler shift of the line. We calculated this Doppler shift of Si\,{\sc iv} at each spatial pixel in the burst for all the available rasters (that are free of cosmic ray hits). One of these Doppler shift maps is displayed in Fig.\,\ref{dopp}(b). Spatially decoupled blue- and redshift regions, not extending beyond the intensity structure itself, are seen in this Doppler map. This spatially decoupled flow pattern is present throughout the lifetime of the UV burst (see the online movie).

\begin{figure}[!htb]
\centering
\includegraphics[width=8.5cm]{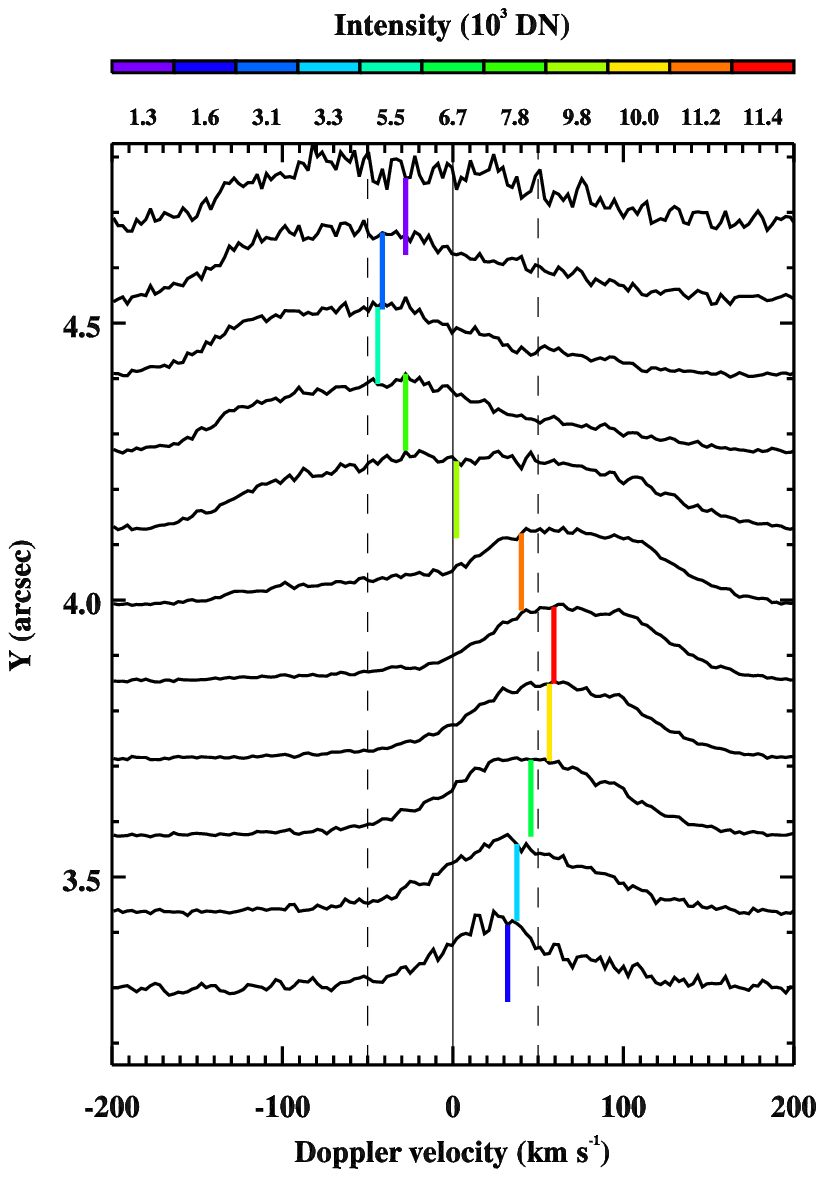}
\caption{Si\,{\sc iv} spectral profiles observed along the vertical line in Fig.\,\ref{dopp}. Each spectrum is normalised to its maximum and shifted in $y$-direction by a constant amount to create this stack plot. The $y$-axis indicates the spatial extent along the vertical line in Fig.\,\ref{dopp}. The coloured vertical bars associated with each spectrum show the position of the intensity-weighted Doppler shift of the line, while the colour of the bar indicates the line intensity, according to the colour bar above the plot. Vertical lines indicate zero and $\pm50$\,km\,s$^{-1}$ Doppler shifts. See Sect.\,\ref{obsv4}.
\label{avgdopp}}
\end{figure}

To clarify the nature of these red- and blueshift patterns in the Doppler maps, we investigated spectral profiles at individual spatial pixels. As a demonstration, we show the Si\,{\sc iv} line profiles observed along the vertical line in Fig.\,\ref{dopp} as a stacked plot in Fig.\,\ref{avgdopp}. The Doppler shifts calculated from the first moments and the profiles as a whole show a swing from red- to blueshift when scanning from south to north of the burst. The flow pattern smoothly changes from redshifts (downflows) to blueshifts (upflows) across a zero line (line with small or zero net Doppler shift) at about $4.2\arcsec$ (see also Fig.\,\ref{dopp}b). The maximum Doppler shift is approximately 50\,km\,s$^{-1}$ in this example. However, the spectral lines have emission at Doppler shifts in excess of $\pm200$\,km\,s$^{-1}$. Beyond the extent of the burst, the spectral profiles are weak and noisy, suggesting that the flows are closely related to the spatial structure of the burst. 

In the first observations of UV bursts with IRIS, \citet{2014Sci...346C.315P} detected a clear double-peaked Si\,{\sc iv} line profile in a single spatial pixel for one burst. While we also observe a few such double-peaked profiles in our data, they constitute only a minority. The majority of pixels show uni-directional bulk Doppler shifts with very broad extensions to the red and blue parts of the spectrum, as shown in Fig.\,\ref{avgdopp}. This suggests that the observed blue- and redshifts are a result of bulk plasma flows or jets in opposite directions.

The Doppler shifts of plasma flows found here exceed the local sound speed (about 50\,km\,s$^{-1}$) and reach a decent fraction of the Alfv\'{e}n speed (a few 100\,km\,s$^{-1}$) in the transition region. This leads to the interpretation that the observed plasma flows are bi-directional jets that are due to magnetic reconnection \citep[e.g.][]{1991JGR....96.9399D,1997Natur.386..811I,2014Sci...346C.315P}. \citet{1997Natur.386..811I} argued that the spatial offset in the Doppler shifts can be interpreted as a magnetic reconnection geometry that is inclined to the line of sight, producing inclined jets. The spatially decoupled flow pattern observed here can be interpreted as inclined reconnection jets. 

In a recent study, \citet{2015ApJ...810...38K} reported a UV burst in the same active region AR 12121, but three days later, on 2014 July 30. As in the present case, the event was associated with a parasitic positive-polarity feature moving away from the sunspot. The authors reported a spatial pattern of blue- and redshifts associated with this UV burst that may be similar to what we report here, although a velocity map such as Fig.\,\ref{dopp}(b) was not presented. This velocity pattern may therefore be a common occurrence in UV bursts associated with moving magnetic features near sunspots. Furthermore, these spatial patterns of reconnection outflows seem to persist for days. This suggests that the UV bursts may be triggered in reconnection geometry governed by the large-scale magnetic field topology that evolves on timescales of days.

The close relation of Doppler shifts to the intensity structure of the burst (Fig.\,\ref{dopp}) suggests that the flows do not extend beyond the location of the burst, as seen in the intensity. There could be three reasons for this: (a) the emission from outflowing jets of plasma is so weak beyond the periphery of the burst that we simply do not see the jets, or (b) the jets are heated or cooled as they propagate outwards, such that the plasma temperature is very different from the Si {iv} formation temperature, or (c) a more likely scenario is that the inclined jets dissipate their energy in the ambient high-density medium, the presence of which is supported by the absence of the O\,{\sc iv} lines at the site of the burst (see Sect.\,\ref{obsv3}).


\section{Magnetic field surrounding the UV burst\label{nlfff}}

In this section we study the evolution of the magnetic field topology responsible for the observed UV burst and place it in the context of the magnetic field distribution on larger spatial scales (several tens of megameters). It is clear from Fig.\,\ref{context} that the UV burst is triggered in between a sunspot and a pore with the same polarity (negative in this case). The spatial distribution of this sunspot and pore, including the neighbouring regions, is shown in Fig.\,\ref{sdo}(a). A closer inspection of the magnetograms shows that a parasitic opposite-polarity magnetic element moving from the sunspot towards the pore is underlying the UV burst (see also Fig.\,\ref{context}). Overlying the burst are densely packed EUV coronal loops rooted in the pore (Fig.\,\ref{sdo}b). The other footpoints of these loops are rooted several megameters away, still in the field of view of Fig.\,\ref{sdo}.

\subsection{Photospheric magnetic structure and reconnection\label{nlfff1}}

We now examine the conditions that led to the onset of the observed UV burst. Recent studies suggest that there is some connection between the transition region UV bursts observed with the IRIS and the H$\alpha$-wing enhancements in the photosphere seen in Ellerman bombs \citep{2015ApJ...812...11V,2016ApJ...824...96T}. UV bursts and Ellerman bombs have both been associated with magnetic reconnection at some height in the solar atmosphere. Ellerman bombs have been suggested to be triggered by an undulating magnetic flux tube that undergoes magnetic reconnection in the photosphere at the sections of the field lines that form a $\cup$-type configuration \citep{2002ApJ...575..506G,2004ApJ...614.1099P}. This configuration is also proposed to explain filamentary structures in sunspot penumbrae \citep{2002Natur.420..390T}. Observations and simulations suggest that these undulating flux tubes are reminiscent of ongoing flux emergence in active regions \citep{2004ApJ...614.1099P,2010ApJ...720..233C}. Evidently, Ellerman bombs are usually observed in emerging active regions at the sites of closely located mixed-polarity fields, and so are the UV bursts \citep{2014Sci...346C.315P,2016MNRAS.463.2190N}. 

From the suggested connection between Ellerman bombs and UV bursts, it seems reasonable at first that the observed burst is a result of magnetic reconnection in an undulatory flux tube. However, not all Ellerman bombs are associated with UV bursts \citep{2016ApJ...824...96T}. It is is also unclear whether a unified magnetic topology exists that could lead to various events, from Ellerman bombs in the photosphere to UV bursts in the transition region. Moreover, we do not see obvious photospheric flux cancellation signatures that could power the observed UV burst. The question of the trigger of the observed UV burst therefore remains open.

\subsection{Data-driven magnetic field extrapolations\label{nlfff2}}

To investigate the trigger of this UV burst, we explored the magnetic field topology by extrapolating the observed magnetic field in the photosphere to greater heights. Here we used the observed line-of-sight magnetograms as a time-dependent lower boundary condition to drive the magnetic field above the photosphere in a series of nonlinear force-free field states. To achieve this, the magnetic induction equation was evolved using a magnetofrictional relaxation technique \citep{1986ApJ...309..383Y}, as implemented by \citet{2000ApJ...539..983V}. Further details of the method are described in \cite{2014ApJ...793..112C}.  

\begin{figure*}[!htb]
\centering
\includegraphics[width=17cm]{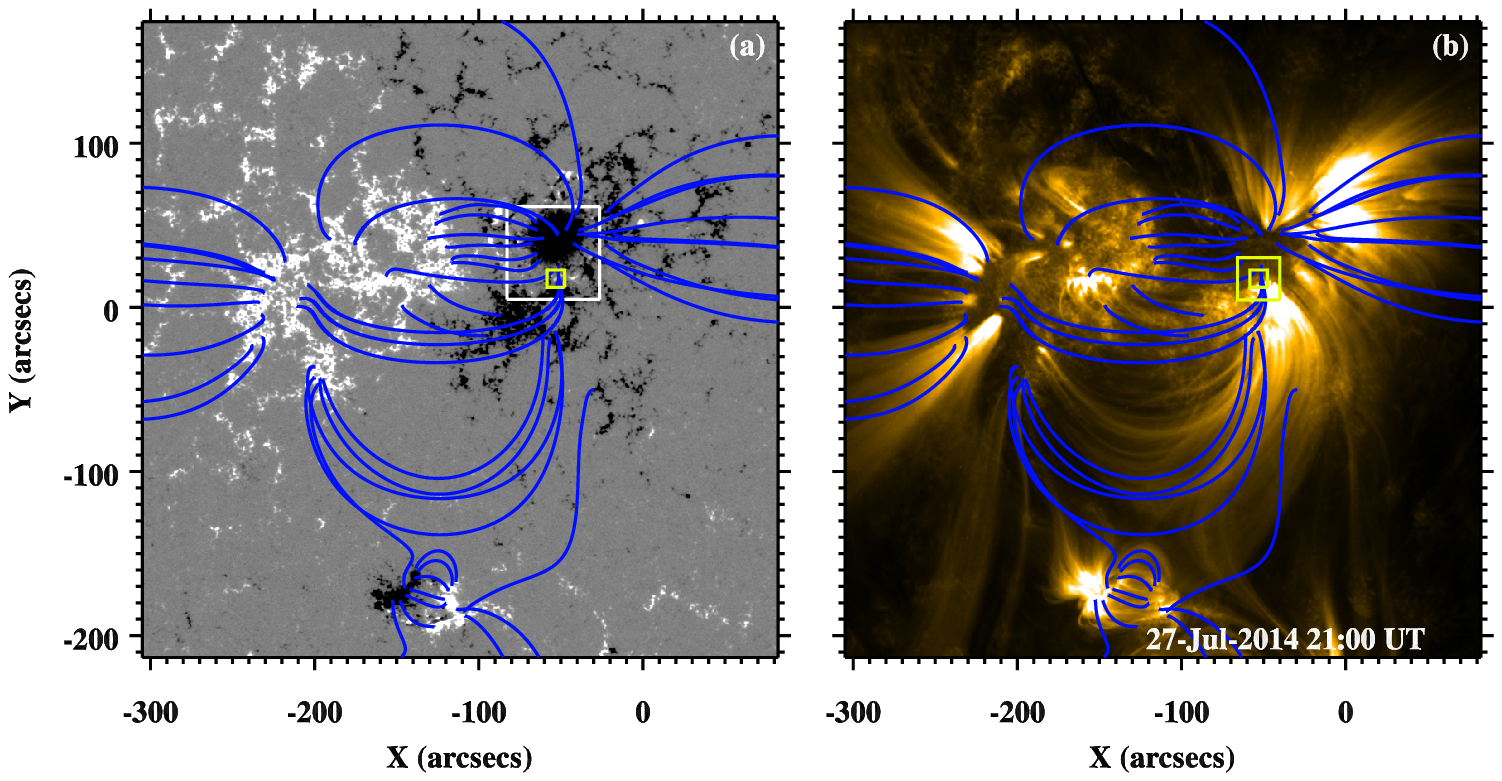}
\caption{Larger magnetic environment surrounding the UV burst. Panel \textbf{a)} shows the map of the SDO/HMI line-of-sight magnetic field saturated at $\pm250$\,G. Panel \textbf{b)} presents
the EUV emission observed with the AIA 171\,\AA\ channel showing several coronal loops. Overlaid in blue on both panels are selected field lines derived from a nonlinear force-free field model evolved with a  magnetofrictional relaxation method. The larger white box in panel \textbf{a)} marks the field of view plotted in Figs.\,\ref{context} and \ref{fan}. The smaller yellow box in the two panels is the location of the UV burst observed with IRIS (yellow box in Fig\,\ref{fan}, see also Fig.\,\ref{mflux}). The larger yellow box in panel \textbf{b)} marks the field of view shown in the top panel of Fig.\,\ref{lcurves}. See Sect.\,\ref{nlfff2}.
\label{sdo}}
\end{figure*}

Overall, the magnetic extrapolations employed here retrieve the coronal field topology reasonably well, as validated by the observed EUV loops and the traced field lines. This is illustrated in Fig.\,\ref{sdo}, where we overplot selected magnetic field lines (blue curves) from the extrapolations on top of the EUV coronal image. The observed loops and traced field lines show that there is a magnetic connectivity between the regions near the sunspot and the extended plage region to the east of the sunspot. Some of these loops originate near the yellow box, the region where the UV burst is triggered. For this reason, it is interesting to connect the dynamics of the UV burst to the coronal loop footpoints in its vicinity. We return to this topic in Sect.\,\ref{cors}.   

\begin{figure*}[!htb]
\centering
\includegraphics[width=17cm]{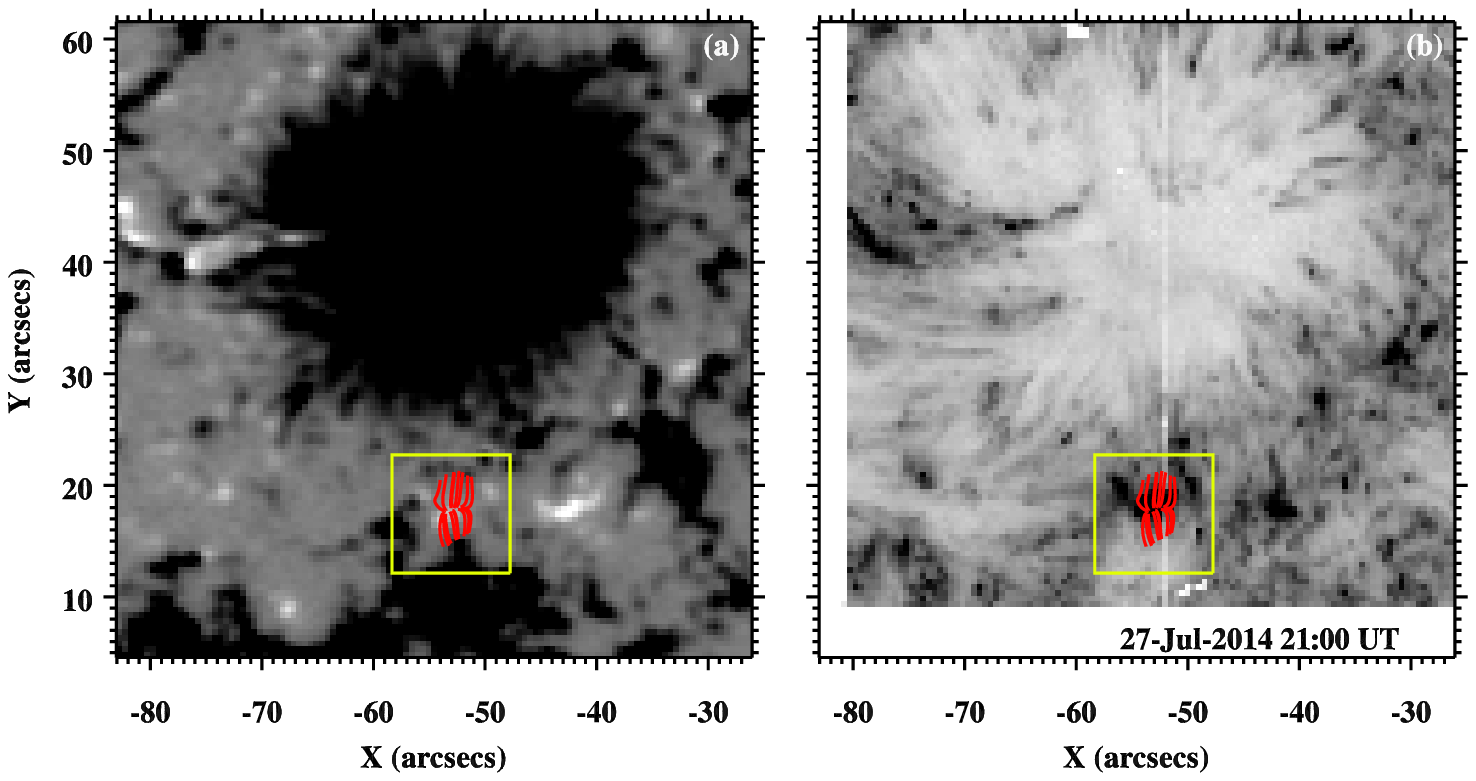}
\caption{Zoom of the magnetic field distribution in the vicinity of the UV burst. \textbf{a)} Map of the line-of-sight magnetic field in the region outlined by the white box in Fig.\,\ref{sdo}. Overplotted in red are selected magnetic field lines that connect the parasitic positive-polarity field with the adjacent dominant negative-polarity field. These closed field lines represent the fan surface. In panel \textbf{b)} the field lines are overplotted on IRIS SJI 1400\,\AA\ image (same as Fig.\,\ref{context}b). The yellow box in the two panels points at the same location as in Fig.\,\ref{sdo}. See Sects.\,\ref{nlfff2} and \ref{nlfff3}.
\label{fan}}
\end{figure*}

The location of this UV burst coincides with the field lines emerging from the opposite parasitic polarity. To illustrate this, in Fig.\,\ref{fan} we zoom into the region outlined by the white box in Fig.\,\ref{sdo}. Here we overplot field lines from the extrapolations on the parasitic positive polarity (left panel) and the UV burst (right panel). These field lines extend on either side to about 2--4\,Mm, with their apexes reaching heights of about 500\,km from the photosphere. Furthermore, these field lines connect the parasitic positive magnetic polarity to the neighbouring magnetic elements closer to the sunspot and the pore of opposite polarity.

\subsection{Fan-spine magnetic topology related to the UV burst\label{nlfff3}}

\begin{figure*}[!htb]
\begin{center}
\resizebox{0.45\hsize}{!}{\includegraphics{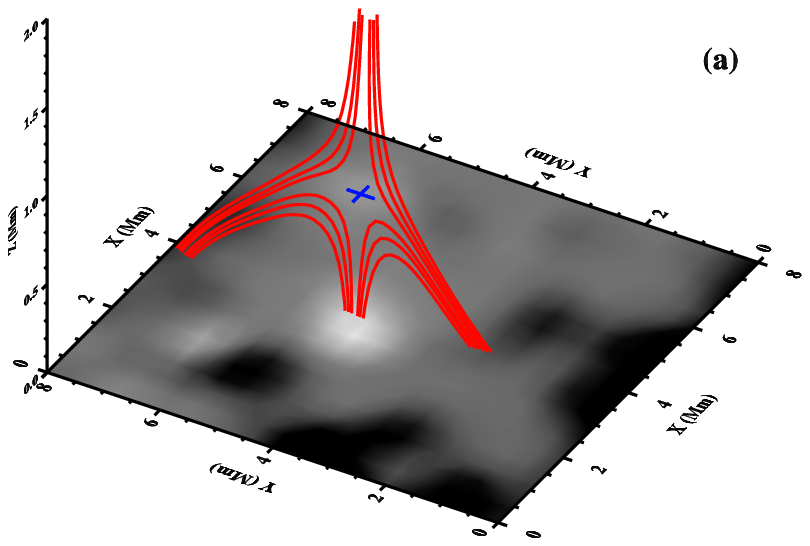}}
\resizebox{0.45\hsize}{!}{\includegraphics{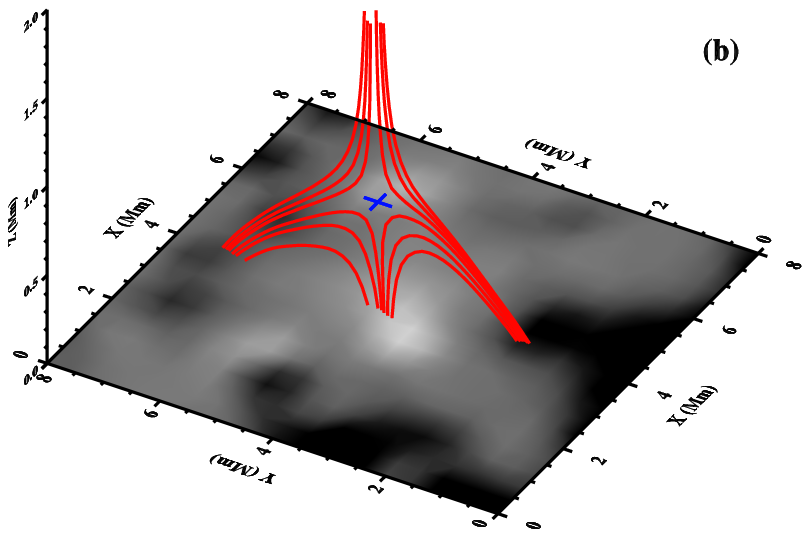}}
\caption{Evolution of the persistent fan-spine magnetic topology (selected field lines plotted in red) and the associated magnetic null (blue $\text{cross}$ , at a height of about 500\,km\ above $z=0$), surrounding the observed UV burst. In each panel, the
gray scale image shows the line-of-sight magnetic field map saturated at $\pm250$\,G. It covers the spatial extent outlined by the yellow box in Fig.\,\ref{fan}. Panel \textbf{a)} is obtained at 20:37 UT, and panel \textbf{b)}  at 21:00 UT. The field lines are plotted only up to a height of 2\,Mm. The $z-$axis is stretched for display purposes. See Sects.\,\ref{nlfff2} and \ref{nlfff3}.
\label{spine}} 
\end{center}
\end{figure*}

At the location of the UV burst, we find a fan-spine magnetic topology that is also found in large-scale flares \citep[e.g.][]{2009ApJ...700..559M,2013ApJ...778..139S}. To show this field line topology near the UV burst, we zoom into the region outlined by the yellow box in Fig.\,\ref{fan} (now covering only the UV burst). This view is shown in Fig.\,\ref{spine}. Here we plot selected field lines that are locally closed (fan surface) and some field lines that recede away and close elsewhere (spine field lines). To demonstrate the persistence of this fan-spine topology of the magnetic field lines, we plot maps obtained 23\,minutes apart in Fig.\,\ref{spine}. From these plots, it is clear that there is a point in the 3D space where the magnetic field vanishes, or in other words, a magnetic null point, which is identified with a $\text{cross}$  in each panel. This system of the magnetic field with a fan-spine topology is prone to magnetic reconnection at the null point following shearing disturbances either in the fan surface or the spine field lines \citep[see e.g.][for an overview of 3D magnetic reconnection regimes]{2011AdSpR..47.1508P}.

\begin{figure}[!htb]
\centering
\includegraphics[width=8.5cm]{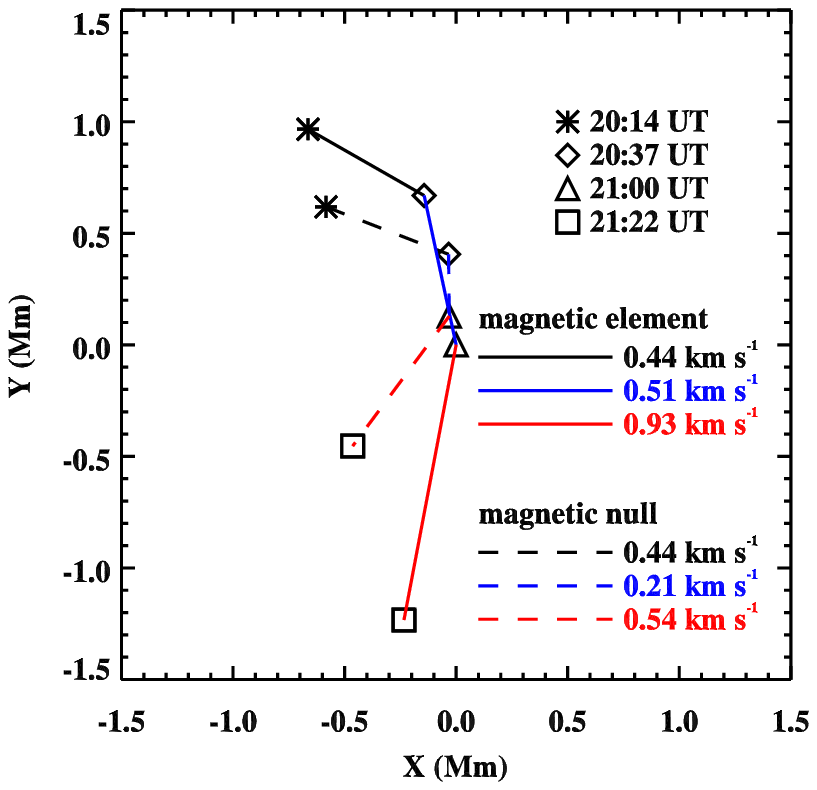}
\caption{Proper motions of the parasitic positive polarity (solid lines) in the photosphere, and the overlying magnetic null (dashed lines) over a period of one hour. The four symbols connecting the lines denote the times at which the positions of the magnetic element and the null are recorded. The lines display the tracks of the respective features. The colours indicate the velocity magnitude with which the respective feature is displaced (i.e. from the recorded positions every 23\,minutes). See Sects.\,\ref{nlfff2} and \ref{nlfff3}.
\label{nullpt}}
\end{figure}

To investigate the nature of shearing disturbances, we now study simple translational motions of the parasitic magnetic element and the overlying null point. To illustrate the persistent long-term motions, we considered four instances of magnetograms, each separated by 30 time steps\footnote{The cadence of HMI magnetograms is 45\,s, which means that 30 time steps amount to a duration of 23\,minutes.}. We identified the positions of the magnetic element in the magnetograms (see Fig.\,\ref{mflux}) and tracked the position of the null point. In Fig.\,\ref{nullpt} we show these tracks separately for the magnetic element (solid lines) and magnetic null (dashed lines). The two features move with speeds slower than 1\,km\,s$^{-1}$. Photospheric magnetic elements typically exhibit proper motions of $\approx1.5$\,km\,s$^{-1}$ \citep[e.g.][and references therein]{2012ApJ...752...48C} and the speeds quoted here should be considered only as average values (since we measure the positions at a 23\,minute cadence). The main result here is that the magnetic element moves faster than the null point on average. The magnetic element is displaced by more than 2\,Mm, while the magnetic null is displaced by only about 1\,Mm during the same time. We interpret this differential motion  between the magnetic element and the null as an indication of magnetic shear between the photosphere and the height of the magnetic null. This means that the fan-spine field lines surrounding the magnetic null are subjected to a shear caused by the motion of the photospheric magnetic element. It is interesting to note that when the magnetic field lines around the null experience maximum shear (between 20:37 UT and 21:00 UT), several episodes of intensity enhancements are also observed in IRIS (see Sect.\,\ref{cors} and Figs.\,\ref{mflux} and \ref{lcurves}). We suggest that these shearing motions caused the formation of a current sheet near the null point, which resulted in the onset of magnetic reconnection and the observed UV burst. This places the formation of the burst in the low chromosphere.

A critical assessment of a variety of nonlinear force-free field (NLFFF) extrapolation algorithms such as the one employed here shows that the accuracy of these algorithms is limited by the quality of the lower boundary (i.e. magnetic maps from observations) and more importantly, by the force-free assumption in the photosphere \citep{2008SoPh..247..269M,2009ApJ...696.1780D}. An additional inaccuracy in energy estimates arises from the basic assumption in all the NLFFF models that plasma-$\beta$ is negligible \citep{2015A&A...584A..68P}. Data-driven 3D magnetohydrodynamic models will overcome many of the limitations in NLFFF models, barring the limitation in the quality of the magnetic field in the lower boundary from observations, and provide a more accurate representation of the magnetic field above the photosphere. However, they are computationally expensive and are beyond the scope of the present work.

To mitigate the problem of the magnetic field line connectivity, we here considered a large enough field of view that completely
covers the main active region along with a neighbouring active region to the south (see Fig.\,\ref{sdo}a). Firstly, the field lines traced in the extrapolations match the observed coronal loops reasonably well. Secondly, the identified magnetic null is stable and the observed burst closely follows it. These two points give confidence in the retrieved fan-spine magnetic topology surrounding the burst.

\subsection{Intermittent nature of the UV burst\label{nlfff4}}

The event presented here can be considered the result of a prolonged magnetic reconnection releasing bursts of energy in the chromosphere and transition region in an episodic fashion with a duration of 5--10\,minutes per burst. The SJI 1400\,\AA\ light curve from the region covering the UV burst is plotted in Figs.\,\ref{mflux} and \ref{lcurves} (red curve) to show this intermittent nature. This time period is longer than classical transition region explosive events that have lifetimes of $\approx60$\,s \citep{1989SoPh..123...41D}. Superimposed on the 5--10\,minutes bursts are also shorter-duration intensity fluctuations with timescales of about 60\,s. 

It will be interesting to see how the reconnection in the fan-spine topology reported here relates to reconnection models of other transition region transients such as explosive events. Recently, \citet{2015ApJ...813...86I} performed numerical simulations of magnetic reconnection leading to explosive events. Starting with an anti-parallel magnetic field geometry, they concluded that the plasmoid instability is a possible scenario to explain the observed line profiles. Using a different approach, \citet{2009ApJ...702....1H} started with a magnetic field including null points and used a wave-induced magnetic reconnection model to explain the properties of explosive events. Their model produced spicule-like jets and spectral signatures similar to those of observed explosive events. Future models and observations will have to show whether a unifying magnetic scenario can be found for all these events.


\section{Coronal signatures\label{cors}}

\begin{figure}[!htb]
\centering
\includegraphics[width=8.5cm]{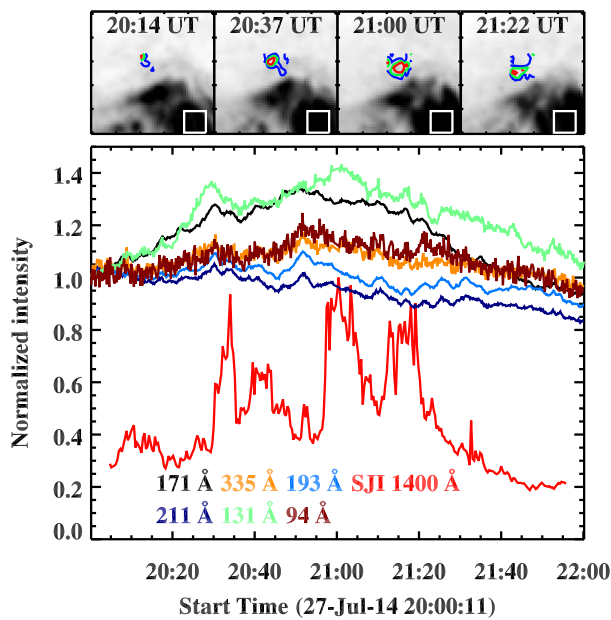}
\caption{Evolution of EUV emission at the footpoints of coronal loops in the vicinity of the UV burst as recorded by AIA and the UV burst as seen in IRIS SJI 1400\,\AA\ images. The top panel shows a series of AIA 171\,\AA\ snapshots (in negative scale) within the larger yellow box plotted in Fig.\,\ref{sdo}(b), which covers an area of 26\arcsec$\times$26\arcsec. The white box in each tile indicates the location of the coronal loop footpoints. The contour lines highlight the location of the UV burst in the same way as shown in Fig.\,\ref{mflux}. In the bottom panel we
plot AIA and IRIS SJI light curves. Here each AIA light curve is the average emission from the region outlined by the white box in the top panel. These AIA light curves are background subtracted and normalised to the respective intensities at their start times. The SJI intensity is normalised to its maximum value. The same SJI light curve is also plotted in Fig.\,\ref{mflux}. See Sect.\,\ref{cors}.
\label{lcurves}}
\end{figure}

The question remains whether the UV burst reported here shows any signatures in coronal emission. From Fig.\,\ref{lcurves} we see that the UV burst has no clear signature in the EUV emission. However, when we plot the EUV emission near the footpoints of coronal loops in the vicinity of the burst, as observed by various AIA channels, we do see some correlation between the SJI 1400\,\AA\ light curve and AIA light curves (bottom panel in Fig.\,\ref{lcurves}). In particular, the channels that are sensitive to the emission at temperatures below 1\,MK (e.g. AIA 171\,\AA\ and 131\,\AA) clearly correspond to the evolution of the UV burst. 

The coronal loop footpoints are at a projected distance of about 5\,Mm away from the burst. The correlation between the AIA 171\,\AA\ and 131\,\AA\ emission with the SJI intensity suggests, however, that these two features, i.e. the UV burst and coronal loops, are connected. The magnetic field extrapolations (Sect.\,\ref{nlfff2} and \ref{nlfff3}) show a fan-spine magnetic topology that supports this remote connection between the burst and the coronal loop footpoints.

In a recent study, \citet{2016A&A...587A..20C} showed that the coronal loops rooted in sunspot umbrae exhibit supersonic downflows as inferred from transition region lines (e.g. Si\,{\sc iv}; O\,{\sc iv}). These downflows result from the suppressed convection and subsequent heating cut-off in umbrae such that the coronal loop can no longer support the plasma. However, the loops rooted in the pore here do not show any signature of supersonic downflows. This indicates that there is some energy supply to the loops at the footpoints that keeps the plasma from cooling and draining.
This energy supply is likely provided by underlying magnetic reconnection.

It is clear that the magnetic topology overlying the UV bursts will regulate its influence at coronal heights. The hot explosions reported by \citet{2014Sci...346C.315P} do not show signatures in the AIA EUV channels, whereas the flaring arch filament events investigated by \citet{2015ApJ...812...11V} do show brief signatures in the 171\,\AA\ and 193\,\AA\ channels. These two studies concentrated on events in the cores of emerging active regions that also host
a strong overlying horizontal magnetic field. In contrast, the event reported here is triggered close to a locally vertical field (pore), and this likely channelled part of the magnetic energy from the reconnection process into the coronal loops. Overall, the energy supply to coronal loops due to parasitic polarity magnetic fields at their footpoints is similar to the scenario proposed by \citet{2017ApJS..229....4C}, who showed that coronal loop footpoints are rooted in a small-scale mixed-polarity magnetic field.


\section{Conclusions\label{cncl}}

We reported observations of a UV burst in an evolved active region. With little flux emergence in its vicinity, the burst was triggered by a moving magnetic feature advected by the moat flow of a sunspot in that active region towards a neighbouring pore of the same magnetic polarity as the spot. The event produced very broad chromospheric and transition region spectral lines with plasma flows in excess of $\pm200$\,km\,s$^{-1}$. The spatial structure of the burst is split into zones of predominantly up- and downflows, suggesting that the plasma jets are inclined to the line of sight. 

Data-driven nonlinear force-free field extrapolations from a time sequence of photospheric magnetograms, evolved with a magnetofrictional relaxation technique, revealed a fan-spine topology surrounding the burst. This field topology is different from the frequently observed undulating flux tubes and bald patches in the cores of active regions. The associated magnetic null point is located at a height of $\approx500$\,km above the photosphere. In addition, we observed that the field lines at the null point are sheared as a result of photospheric motions. Magnetic reconnection is likely in this scenario, and it may be responsible for the series of bursts seen in the IRIS SJI 1400\,\AA\ light curve (Figs.\,\ref{mflux} and \ref{lcurves}). Based on the height of the null point, it is likely that the UV burst has its origin in the low chromosphere.   

The coronal response (in EUV emission) of the observed UV burst is weak. However, we do see evidence of the UV burst in the AIA EUV light curves directly at the loop footpoints. In particular, the light curves from the EUV channels sensitive to plasma emission below 1\,MK correspond well  to the evolution of the UV burst over a duration of one hour. This leads to the interesting proposition that if triggered in a favourable magnetic topology, UV bursts may play a role in the  coronal heating as well.


\begin{acknowledgements}
The authors thank the referee for many useful suggestions. L.P.C. acknowledges useful discussions with Klaus Galsgaard. L.P.C. received funding from the European Union's Horizon 2020 research and innovation programme under the Marie Sk\l{}odowska-Curie grant agreement No.\,707837 and acknowledges previous funding by the Max-Planck-Princeton Center for Plasma Physics. P.R.Y. thanks the staff at the Max Planck Institute for Solar System Research for their kind hospitality, and he acknowledges funding from NASA grant NNX15AF48G. Y.-M.H. acknowledges Max Planck/Princeton Center for Plasma Physics for sponsoring his trip to the Max Planck Institute for Solar System Research, and funding from US NSF grant No. AGS-1460169. The authors thank the ISSI, Bern, for support for the team ``Solar UV bursts -- a new insight to magnetic reconnection''.  IRIS is a NASA small explorer mission developed and operated by LMSAL with mission operations executed at NASA Ames Research center and major contributions to downlink communications funded by ESA and the Norwegian Space Centre. SDO data are courtesy of NASA/SDO and the AIA and HMI science teams. This research has made use of NASA's Astrophysics Data System.

\end{acknowledgements}

\bibpunct{(}{)}{;}{a}{}{,}
\bibliographystyle{aa}

\end{document}